\documentclass[]{spie}  

\usepackage{amsmath}
\usepackage{amsfonts}
\usepackage{amssymb}
\usepackage{graphicx}
\usepackage[caption = false]{subfig}
\usepackage[colorlinks=true, allcolors=blue]{hyperref}
\usepackage{siunitx}
\usepackage{multirow}

\DeclareSIUnit\milliarcsecond{mas}

\title{A Visible-light Lyot Coronagraph for SCExAO/VAMPIRES}

\author[a,*]{Miles Lucas}
\author[a]{Michael Bottom}
\author[b,c,d,e]{Olivier Guyon}
\author[b]{Julien Lozi}
\author[f]{Barnaby Norris}
\author[b]{Vincent Deo}
\author[b,d]{S\'ebastien Vievard}
\author[b]{Kyohoon Ahn}
\author[b,g,h]{Nour Skaf}
\author[f]{Peter Tuthill}
\affil[a]{Institute for Astronomy, University of Hawai'i,  640 N. Aohoku Pl., Hilo, HI 96720, USA}
\affil[b]{Subaru Telescope, National Observatory of Japan, 650 N. Aohoku Pl., Hilo, HI 96720, USA}
\affil[c]{Steward Observatory, University of Arizona, }
\affil[d]{Astrobiology Center, NINS, 2-21-1 Osawa, Mitaka, Tokyo 181-8588, Japan}
\affil[e]{College of Optical Sciences, University of Arizona, 933 N. Cherry Ave., Tucson, AZ 85721, USA}
\affil[f]{Sydney Institute for Astronomy, University of Sydney, Physics Rd., NSW 2006, Australia}
\affil[g]{LESIA, Observatoire de Paris, Univ.~PSL, CNRS, Sorbonne Univ., Univ.~de Paris, 5 pl. Jules Janssen, 92195 Meudon, France}
\affil[h]{Department of Physics and Astronomy, University College London, London, United Kingdom}

\authorinfo{Further author information: (Send correspondence to M.L.)\\M.L.: E-mail: mdlucas@hawaii.edu}

\begin{document}
\maketitle


\begin{abstract}
   We describe the design and initial results from a visible-light Lyot coronagraph for SCExAO/VAMPIRES. The coronagraph is comprised of four hard-edged, partially transmissive focal plane masks with inner working angles of \qtylist{36;55;92;129}{\milliarcsecond}, respectively. The Lyot stop is a reflective, undersized design with a geometric throughput of 65.7\%. Our preliminary on-sky contrast is \num{e-2} at \ang{;;0.1} to \num{e-4} at \ang{;;0.75} for all mask sizes. The coronagraph was deployed in early 2022 and is available for open use.
\end{abstract}




\section{Introduction}\label{sec:intro}

In recent decades exoplanet science has advanced tremendously, with over \num{5000} confirmed planet detections\cite{akeson2013}. These planets have predominantly been discovered and characterized using the indirect methods of transit photometry or radial velocity analysis\cite{perryman2018}. High-contrast imaging, or direct imaging, complements the phase space of exoplanet size, masses, and orbits that these indirect methods probe and is the only way to directly observe exoplanet spectra and orbits. To accomplish the difficult task of imaging a planet around another star, a combination of high-performance instrumentation, like large diameter telescopes and low-noise detectors, observational techniques, like angular differential imaging, and post-processing algorithms must be all be employed.

In the past decade, high-contrast instruments like SPHERE\cite{petit2014}, GPI\cite{macintosh2014}, and SCExAO\cite{jovanovic2015a} have greatly advanced the field of direct imaging. These instruments typically operate in the near-infrared (near-IR), from Y to L band, for two reasons. First, the increased brightness of thermal emission of exoplanets in the near-IR lowers the relative flux ratio between the star and the exoplanet (contrast) and ultimately improves the detection sensitivity to smaller planets. Second, in the near-IR, wavefront control with adaptive optics (AO) is easier than in the visible. This is because the same path length error is a smaller phase error in the near-IR than the visible, simply due to the wavelength difference.

Despite the challenges of high-performance, diffraction-limited imaging at shorter wavelengths, there are significant motivations for extending high-contrast imaging into the visible. First is the detection of reflected light from exoplanets, which is considered the most promising method of imaging terrestrial planets\cite{traub2010}. Second, an important component of improving planet-formation theory is studying the birth of exoplanets in protoplanetary disks, which can be imaged using light reflected from small dust grains suspended in the gas disk. Such observations of protoplanetary disks complement radio imaging of the debris and probe smaller angles, better angular resolutions, and different grain chemistries than near-IR reflected light imaging. Lastly, using H$\alpha$ emission (\qty{656.28}{\nano\meter}) as a tracer for accretion allows direct imaging of matter in-flows onto forming exoplanets\cite{currie2022}, as well as studying matter out-flows in giant stars\cite{norris2020}. Some of these science cases are out of reach of current ground-based telescopes, but it is still important to understand the limitations of the techniques and how they can be applied to future ground and space-based telescopes.

\begin{figure*}
   \centering
   \includegraphics[width=\textwidth]{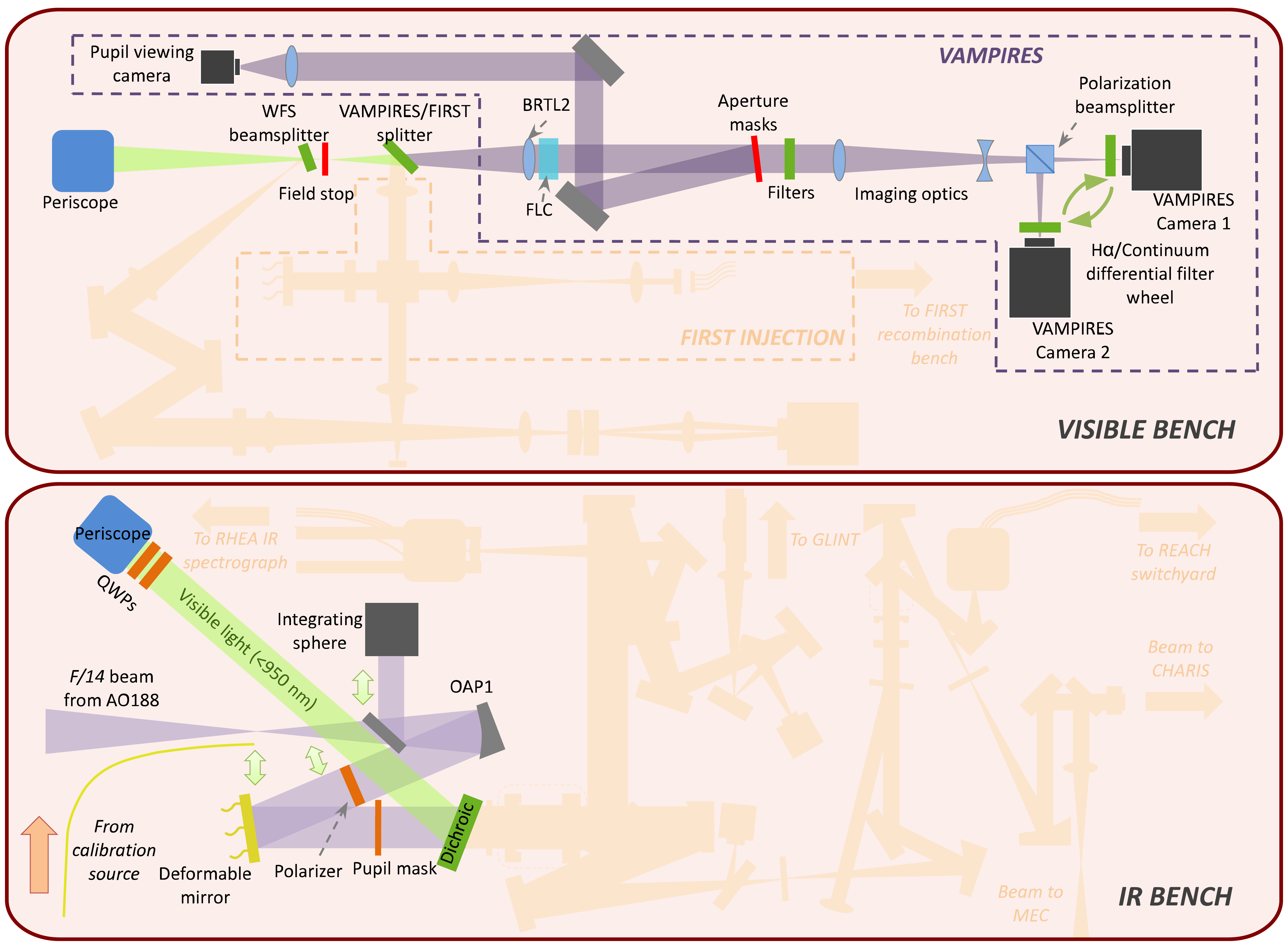}
   \caption{Schematic of SCExAO and VAMPIRES. The first two levels of SCExAO are shown in each box, with all unrelated components faded out. A dashed line outlines VAMPIRES on the visible bench. The focal plane masks for the coronagraph are installed on the visible bench in the field stop (red rectangle, left) and the Lyot stop is installed in the pupil mask wheel (red rectangle, middle).}\label{fig:bench}
\end{figure*}

The Visible Aperture-Masking Polarimetric Imager/Interferometer for Resolving Exoplanetary Signatures (VAMPIRES)\cite{norris2015} is an optical dual-beam imager for the Subaru Coronagraphic Extreme Adaptive Optics (SCExAO) instrument on the Subaru telescope. A schematic of SCExAO/VAMPIRES is shown in \autoref{fig:bench}. VAMPIRES operates between \qtyrange{600}{800}{\nano\meter} with capabilities for polarimetric imaging\cite{norris2020}, H$\alpha$ imaging\cite{uyama2020}, and focal plane wavefront sensing\cite{vievard2020,vievard2022,deo2022}. VAMPIRES also has a suite of non-redundant sparse aperture masks for sub-diffraction-limited interferometric imaging. For this work, though, we will focus on the traditional and polarimetric imaging modes.

VAMPIRES currently uses two electron-multiplying (EM) CCDs (model Andor iXon 897) for low read noise with framerates from \qtyrange{1}{100}{\hertz}. While these detectors perform very well in the photon-limited regime, the strong non-linearities and compressed dynamic range from electron multiplication makes imaging bright targets ($m_I < 7$) difficult. To address this problem, we have designed and deployed a classic Lyot coronagraph (CLC), which greatly attenuates the stellar point spread function (PSF), allowing the use of high EM gain without saturation. This is the primary motivation for building a coronagraph for VAMPIRES- avoiding saturation is key for CCD imaging, but this limits the off-axis signal-to-noise ratio (S/N). By attenuating the on-axis light, the off-axis signal can be increased without saturation from the stellar PSF, ultimately improving the S/N.

Furthermore, coronagraphy improves the detection sensitivity of circumstellar regions by controlling the stellar diffraction pattern. In other words, the coronagraph does not simply mask the central region of the image but rather attenuates the effects of the stellar PSF throughout the entire field of view (FOV). The diffraction control of a coronagraph is highly dependent on the quality of the incoming wavefront, though, necessitating the use of AO for high-contrast imaging. When combined with extreme AO, such as provided by SCExAO, a coronagraph can enable detections of sources many orders of magnitude fainter than their host stars ($\sim$\num{e-6})\cite{guyon2018}.

This report details our design and deployment of a visible-light coronagraph for VAMPIRES. In \autoref{sec:methods} we detail our modeling and simulation of the coronagraph and the construction of the optics. In \autoref{sec:tests} we describe our characterization of the coronagraph, including throughput and inner working angle. Finally, in \autoref{sec:results} we discuss the performance of the coronagraph with the calibration light source and on-sky.

\section{Methods}\label{sec:methods}

\subsection{Coronagraph Design}\label{sec:design}

\begin{figure*}
   \centering
   \includegraphics[width=\textwidth]{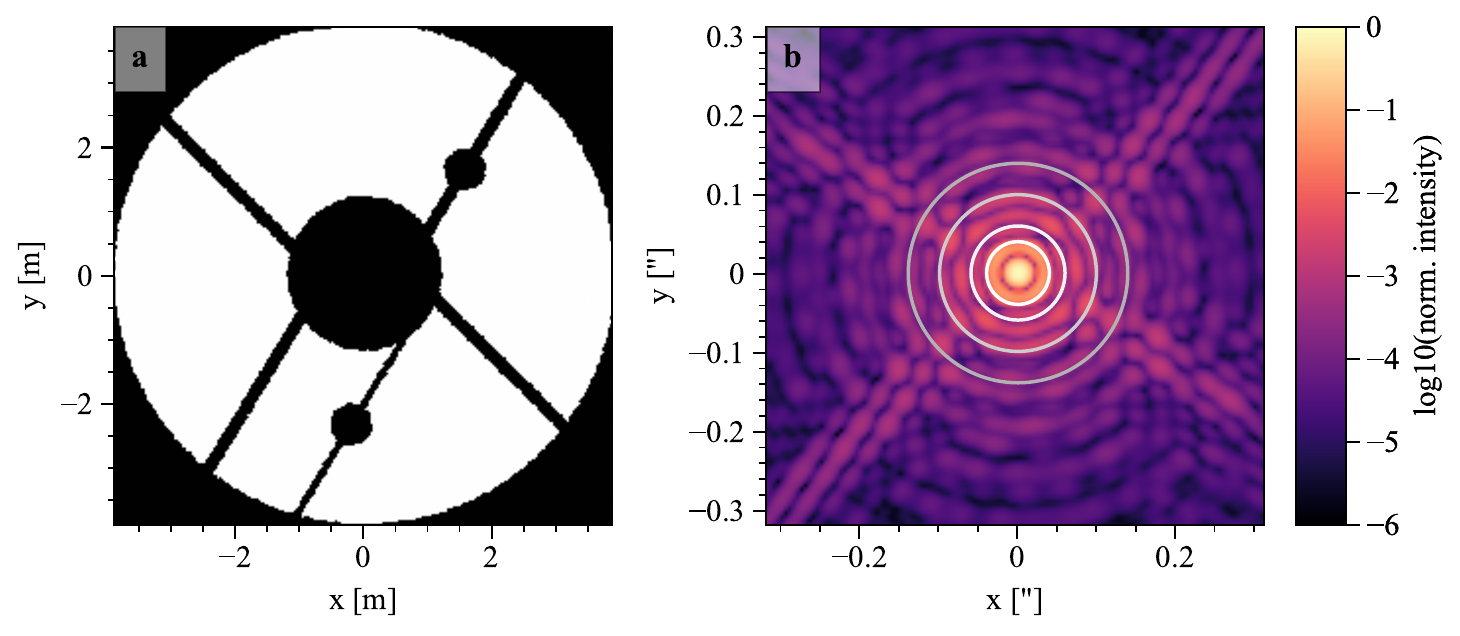}
   \caption{(a) The SCExAO pupil, which is slightly undersized compared to the Subaru aperture and contains additional features. The aperture diameter is 95\% of the Subaru clear aperture; effectively \qty{7.79}{\meter}. The central obstruction is 31\% of this diameter. The two additional circles are masks for inactive deformable mirror actuators, one of which lies on a spider segment, while the other requires an additional spider. (b) The VAMPIRES PSF simulated using \texttt{HCIPy}. The four grayscale circles correspond to the coronagraph focal plane mask sizes.}\label{fig:pupil}
\end{figure*}

We used the open-source python package \texttt{HCIPy}\cite{por2018} for simulating our coronagraph design with Fourier optics. We defined our wavefronts on a rasterized image of the SCExAO pupil sampled on a 256x256 pixel grid (\autoref{fig:pupil}a). Initially, we do not introduce any wavefront errors. We used existing pupil masks to determine the beam diameter in VAMPIRES precisely (\qty{7.03}{\milli\meter}), which corresponds to a focal ratio of F/28.4 for the coronagraph. With \texttt{HCIPy}, we use a Fraunhofer propagator with this focal ratio to form the intermediate focal plane, where the focal plane mask will be inserted. VAMPIRES has a FOV of \ang{;;3}$\times$\ang{;;3}, however, for our exploratory simulations we only calculated the first 16 airy rings, or $\sim$\qty{38}{\milliarcsecond} (\autoref{fig:pupil}b). The focal plane masks are hard-edged circles and were modeled using a circular apodizer with either 0\% or 0.1\% transmission. We designed four focal plane masks with radii of \qtylist{38.6;57.9;96.5;135}{\micro\meter}, respectively. These radii correspond to \numlist{2;3;5;7} $\lambda/D$ at \qty{750}{\nano\meter}, or \qtylist{35.9;53.9;89.8;126}{\milliarcsecond}. The focal plane after the mask was propagated to the next pupil plane, where the Lyot stop apodized the wavefront.

The Lyot stop was modeled using a series of circular and rectangular apodizers with 0\% transmission. The stop design follows the shape of the SCExAO pupil with an undersized diameter and oversized obstructions. The outer diameter is 90\% the clear aperture diameter, the inner obstruction is 45\% the clear aperture diameter, the secondary support struts are 200\% their original width, and the bad deformable mirror actuators are 150\% oversized. The ratio of areas of the Lyot stop and the SCExAO pupil is 63.7\%, and will theoretically widen the diffraction limit by 10\%. This final wavefront was propagated to the detector focal plane, forming the post-coronagraphic point-spread function.

We repeated the above process for a series of wavelengths to approximate a broadband PSF. We summed the images produced with 10 wavefronts from \qtyrange{725}{775}{\nano\meter} to mimic the ``750-50'' filter (\qty{750}{\nano\meter} central wavelength with \qty{50}{\nano\meter} uniform bandpass). We chose this filter because at these wavelengths VAMPIRES has the highest throughput and best polarimetric efficiency. As an additional step, we generated many datasets with random tip and tilt errors sampled from a bivariate Gaussian with \qty{10}{\milliarcsecond} root-mean-square (RMS) jitter to test the coronagraph's resiliency to tip-tilt jitter. SCExAO typically has a higher RMS tip-tilt error than \qty{10}{\milliarcsecond}, which is closer to the 90th-percentile of on-sky performance. However, since VAMPIRES often operates in high-framerate lucky imaging mode, the effects of tip-tilt can be partially mitigated: the apparent movement of the focal plane mask will blur the edge, degrading sensitivity close in, but the rest of the FOV will be sharpened.

\begin{figure*}
   \centering
   \includegraphics[width=0.95\textwidth]{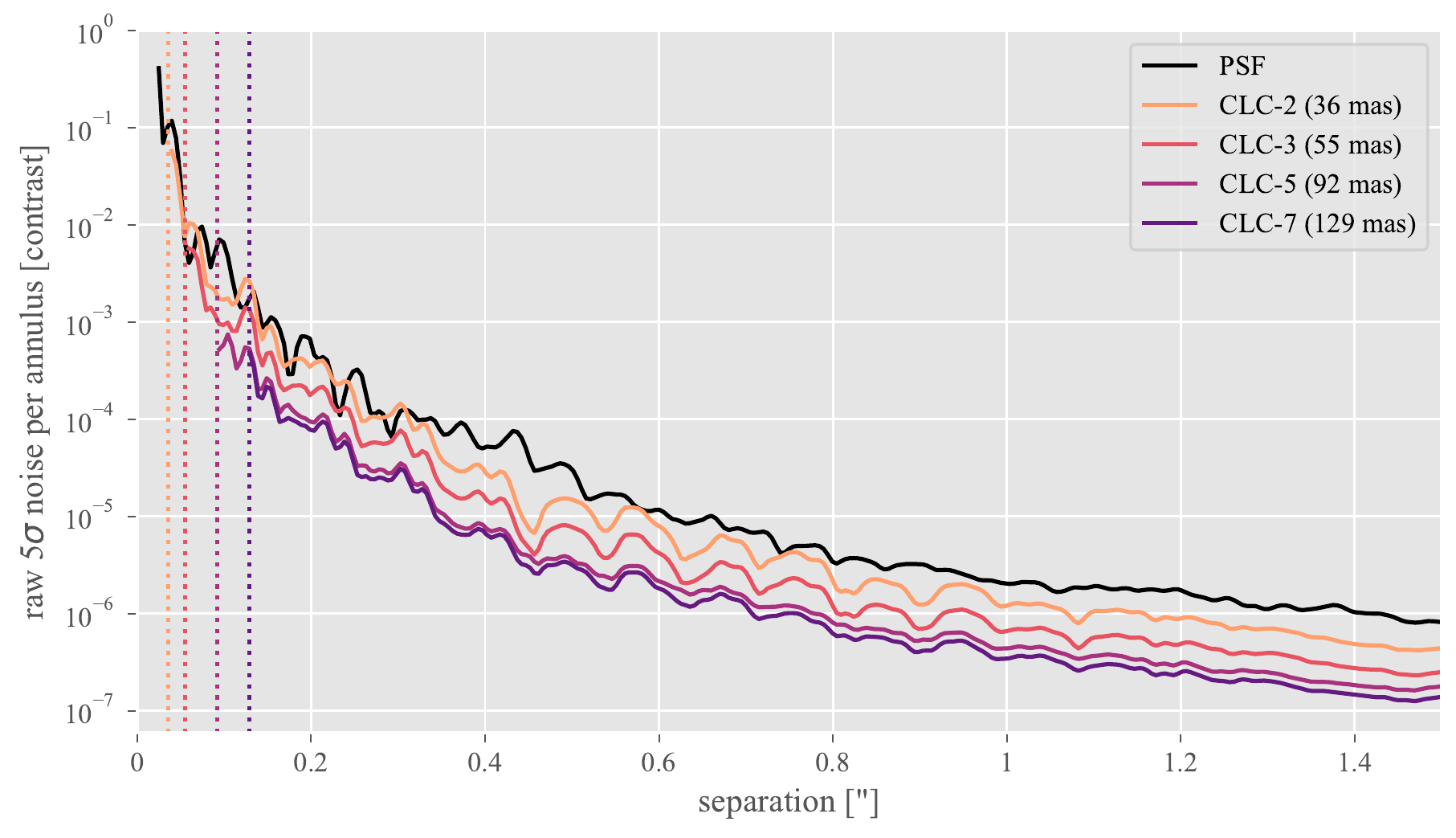}
   \caption{Simulated raw 5$\sigma$ contrast curves for an unocculted PSF (black line) and for each focal plane mask size (orange to purple lines). These curves are calculated from images simulated with \texttt{HCIPy}. The mask inner working angles are marked by vertical dotted lines}\label{fig:sim-contrast}
\end{figure*}

We evaluated the performance and quality of our simulations from the raw contrast curves of each mask as shown in \autoref{fig:sim-contrast}. These curves measure the noise in concentric annuli around the PSF. Each annulus is one full-width at half-maximum (FWHM) wide, and the noise is calculated using the biweight scale statistic, which is robust to outliers such as cosmic rays. The noise estimate also takes into account small-sample statistics\cite{mawet2014}, which bias the noise measured in annuli close to the PSF. We take the radial profile of this noise and normalize it to the non-coronagraphic PSF flux to determine the raw contrast, where we then apply a 5$\sigma$ detection threshold. For clarity, we describe all our contrast curves as ``raw'' since we never do PSF subtraction in this work. To create these contrast curves We used the open-source package \texttt{ADI.jl}\cite{lucas2020}.

From the raw contrast curves, we found that the CLC-2 mask does not attenuate the PSF well, especially at close angles ($<$\ang{;;0.4}). The CLC-3 mask only offers a factor of a few improved contrast compared to the unocculted PSF. We found no significant performance differences between the CLC-5 and CLC-7 masks, both showing an improvement of an order of magnitude of sensitivity throughout the FOV. The typical read noise and non-common path wavefront errors in VAMPIRES are expected to ultimately limit our contrast. As we will describe in \autoref{sec:install}, the focal plane masks are diced from a single piece of glass producing four substrates. We constructed all four mask sizes with two variations each: one partially transmissive (0.1\%) and the other fully opaque.

\subsection{Construction and Installation}\label{sec:install}

We used a metallic vapor deposition process (\url{https://opto-line.com/}) for applying the focal plane mask and Lyot stop patterns to our substrates. The metals are deposited onto a glass substrate with micrometer-precision, giving us great flexibility and precision in our design patterns and the thickness of the material. One downside of this approach is that the glass focal plane mask shifts the focus since it is located in a converging beam. In VAMPIRES, there is a custom optics chassis mounted in the focal plane with \qty{8}{\milli\meter} $\times$ \qty{8}{\milli\meter} slots, so we arranged for four masks to be diced from a single \qty{30}{\milli\meter} anti-reflection (AR) coated optical flat (\href{https://www.edmundoptics.com/p/30mm-dia-4mm-thick-nir-i-coated-lambda10-fused-silica-window/27562/}{Edmund Optics \#84-466}), making sure each segment was within the clear aperture. The focal plane mask dots were deposited using chromium with thicknesses of \qtylist{110;300}{\nano\meter}. The thickness was determined from the optical penetration depth for the desired transmission
\begin{equation}
    \hat{s}(\lambda) = -\frac{\lambda}{4\pi\tilde{k}(\lambda)}\ln{\frac{I(\lambda)}{I_0(\lambda)}}
    \label{eqn:throughput}
\end{equation}
where $\hat{s}$ is the thickness, $\lambda$ is the wavelength (tested across the 600-800 nm range for VAMPIRES), $\tilde{k}$ is the extinction coefficient, and $I/I_0$ is the relative intensity for transmitted light. The thicknesses of \qtylist{110;300}{\nano\meter} correspond to transmissions of \num{e-3} and \num{e-8}, respectively, matching our design goals of 0.1\% and 0\%.

The Lyot stop pattern was deposited onto a \qty{25}{\milli\meter} AR-coated optical flat (\href{https://www.edmundoptics.com/p/25mm-dia-3mm-thick-nir-i-coated-lambda10-fused-silica-window/27561/}{Edmund Optics \#84-465}) with no further processing. We chose a reflective gold coating that allows easy mask alignment with VAMPIRES' pupil camera. Following advice from our vendor, this gold layer was deposited on top of a layer of chromium, which adheres better to AR coatings than gold. We used \autoref{eqn:throughput} to determine a combined thickness of \qty{285}{\nano\meter} for \num{e-8} transmission. The final thickness of the chromium and gold deposit was specified as \qty{300}{\nano\meter}.

\begin{figure*}
   \centering
   \subfloat{\includegraphics[width=2.5in]{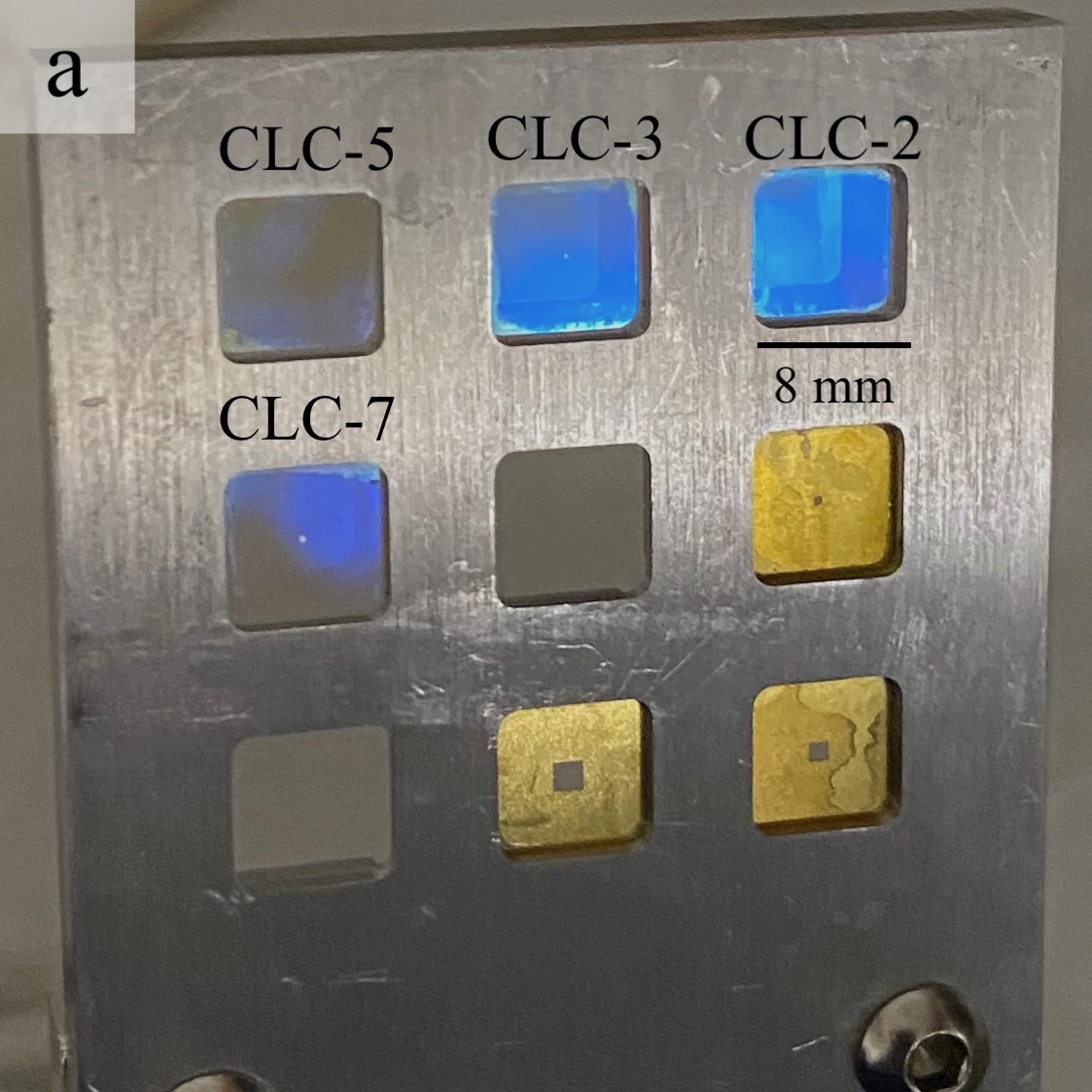}}\hspace{0.5in}
   \subfloat{\includegraphics[width=2.5in]{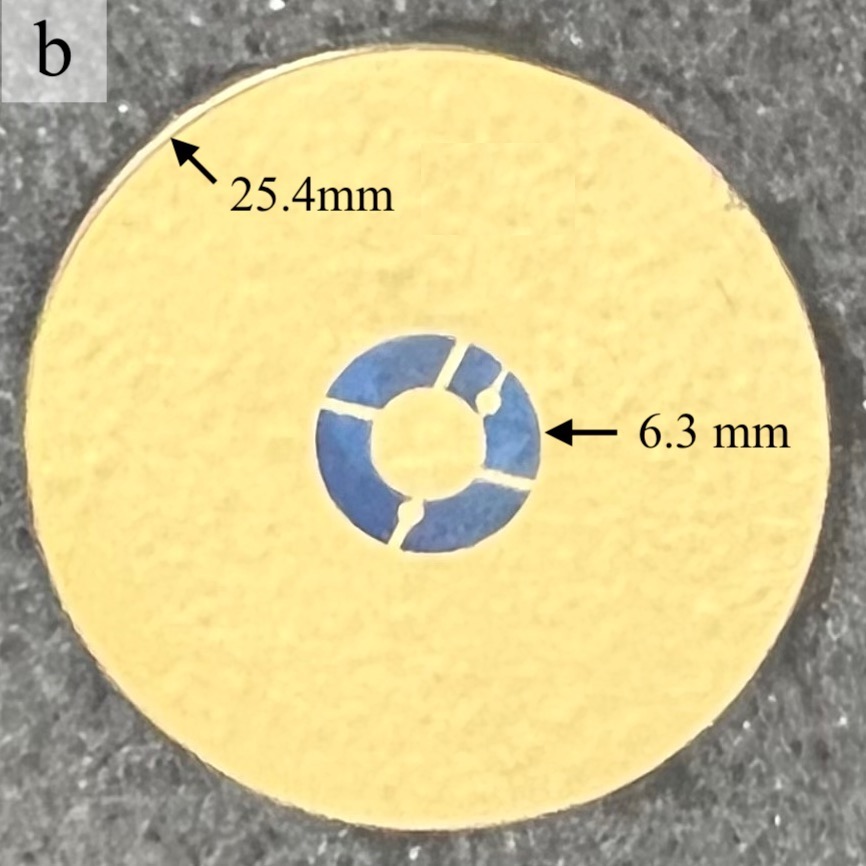}}
   \caption{(a) The four focal plane masks mounted in the custom-machined focal plane mount for VAMPIRES. The size of each slot is \qty{8}{\milli\meter} $\times$ \qty{8}{\milli\meter}. (b) The undersized Lyot stop with its reflective gold coating. The diameter of the glass substrate and stop outer diameter are marked.}\label{fig:optics}
\end{figure*}

\section{Operation and Characterization}\label{sec:tests}

Both the focal plane masks and Lyot stop are installed in mounts with precise motor controllers for alignment. The Lyot stop is aligned using the reflected light in a pupil alignment camera on VAMPIRES. We found that the Lyot stop stays consistently aligned in-between instrument cranings, meaning that no adjustments were necessary for months. The focal plane masks take around a minute to manually align and focus and need to be realigned for each target.

\subsection{Artificial Calibration Speckles (``Astrogrid'')}

\begin{figure*}
   \centering
   \includegraphics[width=\textwidth]{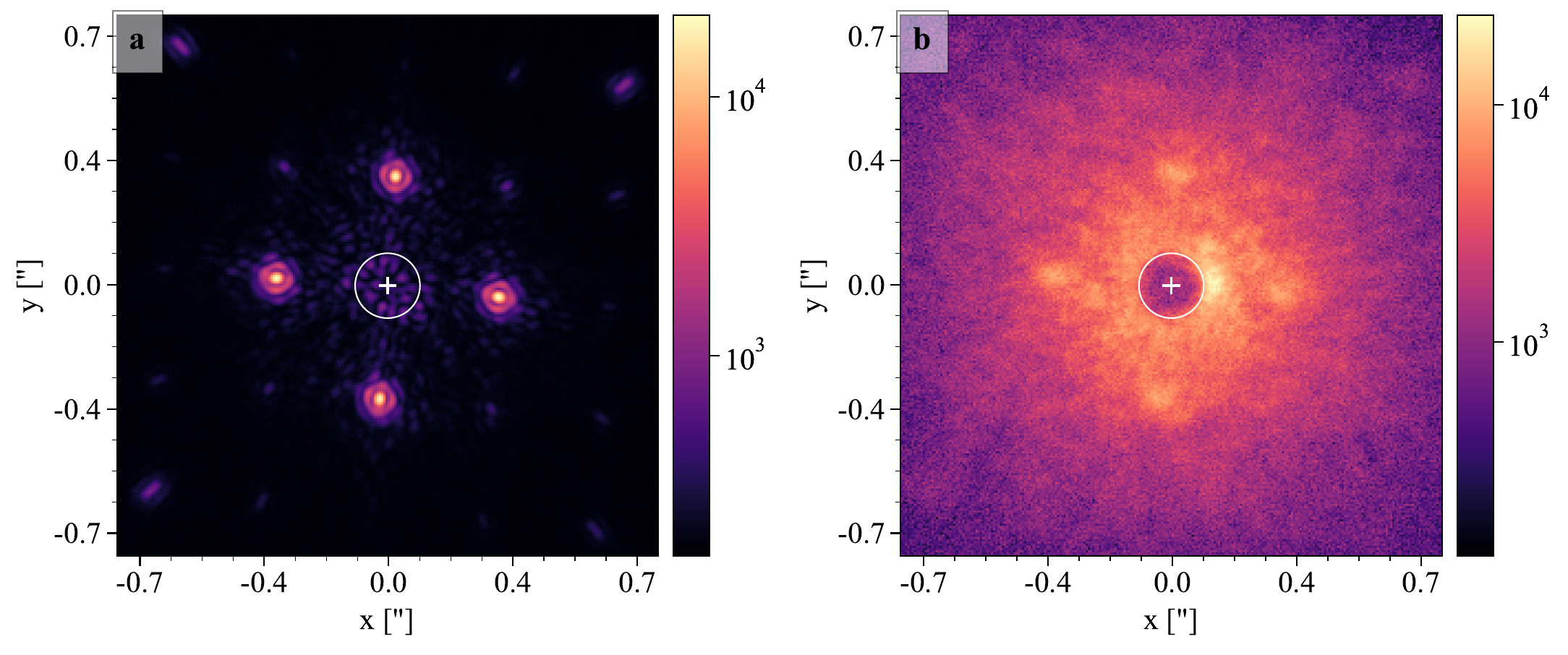}
   \caption{Coronagraphic images using the CLC-5 mask (\qty{92}{\milliarcsecond} IWA, denoted with white circle) with \qty{50}{\nano\meter} ``astrogrid'' calibration speckles. (a) A single frame of testbed data using the internal calibration source through the 750-50 filter with 110 EM gain, and \qty{20}{\milli\second} exposure length. (b) On-sky data of HIP 56083 from May 25, 2022 in poor seeing ($\sim$\ang{;;1.6}), using the 750-50 filter, 300 EM gain, and \qty{200}{\milli\second} exposure length.}\label{fig:satellite-spots}
\end{figure*}

A common problem with coronagraphic imaging is accurately determining the flux and position of the host star while behind the focal plane mask. Even though the focal plane masks in our coronagraph are transmissive, there is not enough signal to precisely measure the photometry or astrometry of the star. Rather, this transmission helps with initial alignment when the Lyot stop is removed allowing a much higher S/N through the mask. Instead, we utilize a technique that creates artificial PSFs throughout the FOV with known photometry and astrometry. This technique, called ``astrogrid''\cite{sahoo2020}, creates ``satellite spots'' by adding sinusoidal patterns to the deformable mirror (DM). These sinusoids create two opposing copies of the PSF, where the sinusoid frequency defines the separation and the amplitude defines the relative brightness. The sinusoid phase is modulated in time to remove coherence between the astrogrid and the underlying speckles. The astrogrid is applied in a cross pattern so that the center of the star can be estimated (\autoref{fig:satellite-spots}). The astrogrid method has been successfully demonstrated on SCExAO/CHARIS with astrometric precision $\sim$\qty{1.7}{\milliarcsecond} and photometric precision $\sim$0.3\%\cite{currie2020}.

\subsection{Coronagraphic Characterization}

We characterized our coronagraph by measuring the Lyot stop throughput, focal plane mask inner working angle (IWA), and focal plane mask transmission. The Lyot stop throughput is proportional to the unobstructed area of the mask compared to the pupil. From our CAD models, we estimated the throughput as 63.7\% using the ratio of unobstructed areas between the Lyot stop and the SCExAO pupil. We measured the throughput using the pupil-imaging camera with the calibration light source, comparing the total intensity measured with a mirror in the pupil wheel (100\% of the incident light) versus the Lyot stop (blocked light only). We calculated a geometric throughput of 65.7\%, which is close to the expected value.

\begin{figure*}
   \centering
   \includegraphics[height=4in]{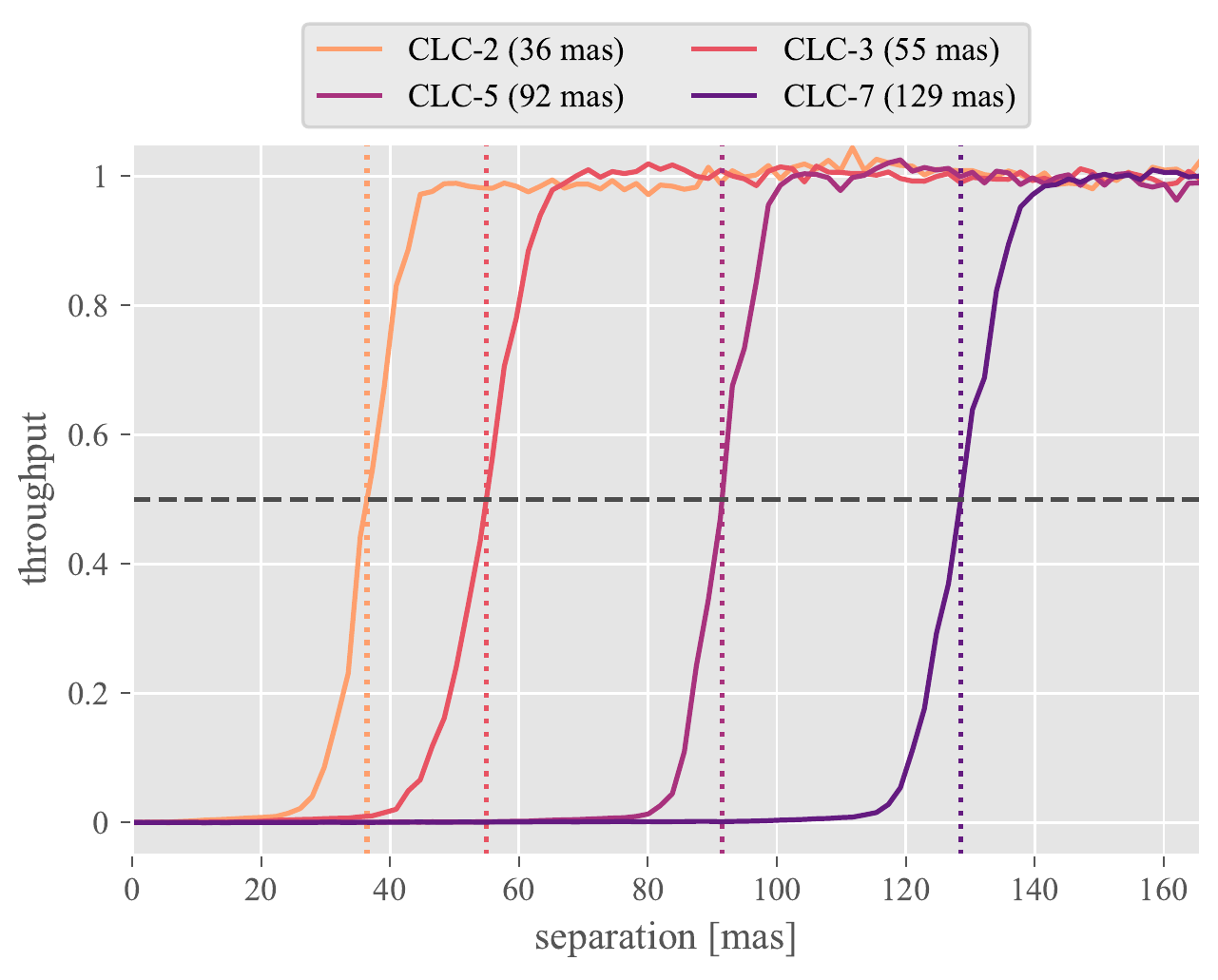}
   \caption{Off-axis coronagraphic throughput for each focal plane mask. The flux throughput is normalized between 0 and 1. 50\% throughput is marked with a black dashed line and the inner working angles are denoted by the orange to purple dotted lines.}\label{fig:throughput}
\end{figure*}

To calculate the IWA we measured the coronagraphic throughput for each mask by gradually moving the focal plane mask off-axis from the calibration source, taking images at each step. The fluxes from all of the images were normalized between 0 for the on-axis flux, and 1 for the maximum off-axis flux (\autoref{fig:throughput}). We interpolated the IWA using the 50\% throughput point from each curve (\qtylist{36;55;92;129}{\milliarcsecond}, respectively). These IWAs are in good agreement with the radii of the masks, as shown in \autoref{tab:coronagraph}.

Lastly, we verified the transmission of the focal plane masks by taking data without the Lyot stop inserted. By removing the Lyot stop, we no longer reject the light blocked by the mask. We measured the total flux behind the masks and compared that to the total flux in the same radius around the unobstructed PSF. Our masks were designed for 0.1\% transmission and the average transmission measured with the calibration source is 0.06\%.

\begin{table}
   \centering
   \caption{Coronagraph parameters for all four focal plane masks.}\label{tab:coronagraph}
   \begin{tabular}{ccccc}
      \hline
      \multirow{2}{*}{Name} & \multicolumn{2}{c}{Mask radius} & IWA &\multirow{2}{*}{Lyot throughput} \\
      & (\unit{\micro\meter}) & (\unit{\milliarcsecond}) & (\unit{\milliarcsecond}) & \\
      \hline\hline
      CLC-2 & 38.6 & 35.9 & 36 & 65.7\% \\
      CLC-3 & 57.9 & 53.9 & 55 & 65.7\% \\
      CLC-5 & 96.5 & 89.8 & 92 & 65.7\% \\
      CLC-7 & 135 & 126 & 129 & 65.7\% \\
      \hline
   \end{tabular}
\end{table}

\section{Results}\label{sec:results}

\subsection{Testbed results}\label{sec:testbed}

Our first tests use the calibration source on SCExAO for determining the contrast of the coronagraph for each focal plane mask. For these tests we used the 750-50 filter with the astrogrid set to \qty{50}{\nano\meter} amplitude and a spatial frequency of 15.5 $\lambda/D$ (\qty{308}{\milliarcsecond}).

First, we took non-coronagraphic data allowing us to characterize the photometry and astrometry of the astrogrid speckles. For data reduction, we median-combined a series of 100 images and centered the frame using the satellite spots. For the astrogrid calibration, we used a hierarchical model of Moffat functions with shared shape and amplitude parameters. Around each PSF we measured the photometry using circular apertures with background subtraction from annular apertures. The relative photometry of the \qty{50}{\nano\meter} amplitude astrogrid was $10^{-1.52}$.

Then we inserted the focal plane mask and Lyot stop and refocused. We median-combined a series of 100 images and centered using the satellite spots. We fit each satellite spot using a hierarchical model of Moffat functions, again. We measured the photometry of each satellite spot using circular apertures with an annulus for background subtraction. The stellar flux was estimated from the median flux of the four satellite spots scaled by the relative photometry fit above ($10^{-1.52}$).

Lastly, we removed the focal plane mask and Lyot stop without adjusting the source brightness or exposure settings. It was paramount that we used 0 EM gain because our detectors saturated heavily without the coronagraph. We took another set of 100 frames, median-combined them, and centered using the satellite spots. All saturated pixels were masked before calculating the contrast, which was normalized using the source flux measured in the coronagraphic images because the satellite spots were partially saturated.

From here we created raw contrast curves the same way as in \autoref{sec:design}, calculating the 5$\sigma$ detection limits in concentric annuli one FWHM wide (accounting for small-sample statistics\cite{mawet2014}) without PSF subtraction. Because we are not performing any kind of differential imaging, the satellite spots heavily bias the contrast, so we mask them out during the noise calculations. We also calculated the background noise from our dark frames and converted it to contrast, which was $\sim$\num{3e-6} (5$\sigma$).

\begin{figure*}
   \centering
   \includegraphics[width=0.9\textwidth]{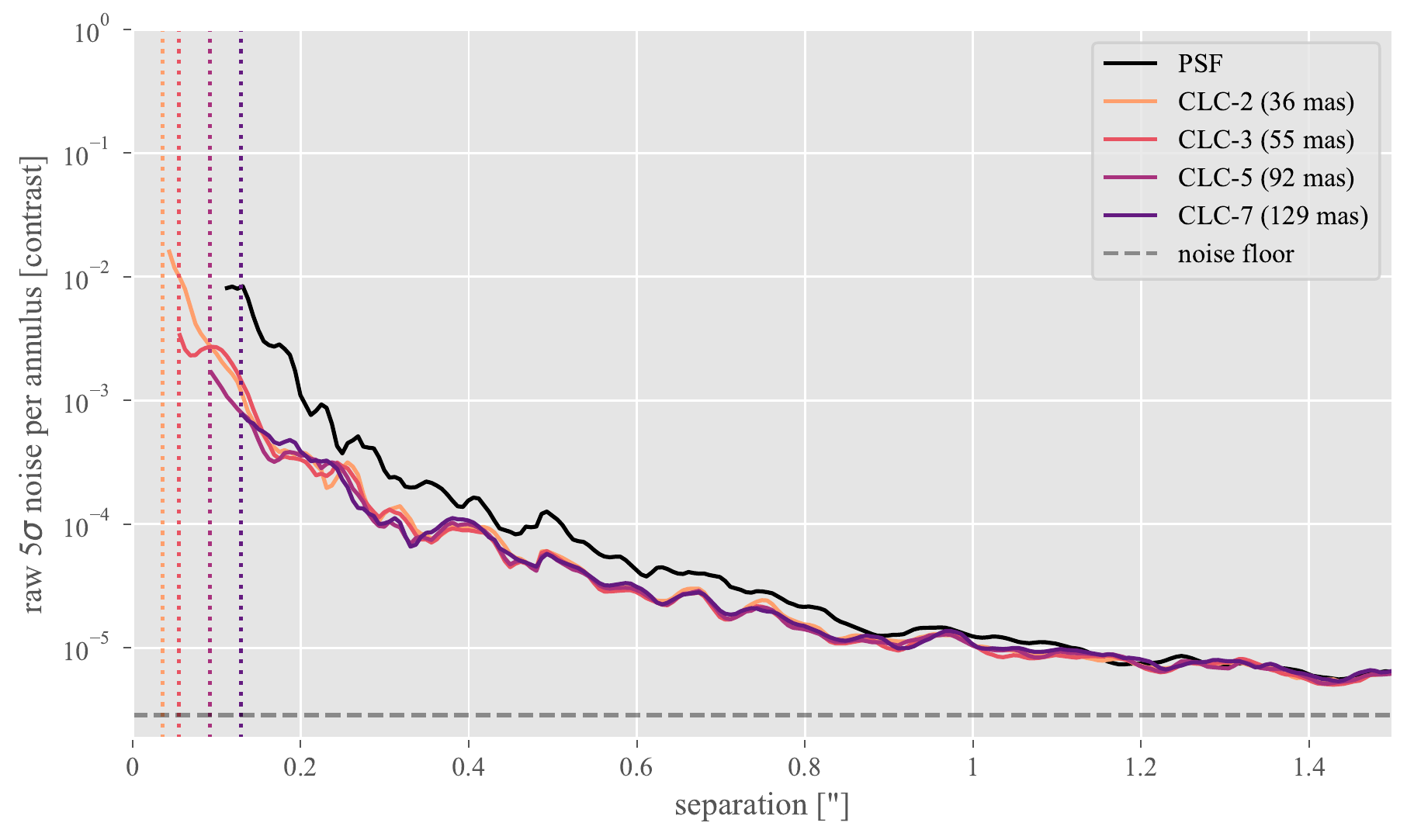}
   \caption{Raw 5$\sigma$ contrast curves measured from \qty{2}{\second} (\num{100} frames) of testbed data. These curves account for small-sample statistics and have no PSF subtraction. The black curve is the non-coronagraphic contrast measured from a saturated PSF (which has been masked out), while the orange through purple curves correspond to each focal plane mask. The IWA for each curve is marked with a dotted vertical line. The astrogrid speckles are also masked out.}\label{fig:contrast}
\end{figure*}

From the raw contrast curves (\autoref{fig:contrast}), we note similar performance between all the mask sizes. At \ang{;;0.1}, the coronagraph achieves a raw contrast of $\sim$ \num{e-3}, and background-limited contrast beyond $\sim$\ang{;;1} of $\sim$\num{e-5}. Throughout the entire FOV, the coronagraphs attenuate the light by \numrange{0.2}{1} orders of magnitude. We expect the diffraction control relative to the unocculted PSF to be consistent throughout the FOV until the noise limit is reached, as in \autoref{fig:sim-contrast}. Here we see very little diffraction control beyond \ang{;;1}, suggesting residual wavefront errors are dominating the noise.

These bench contrast curves represent an ideal observing scenario (i.e., high S/N, no seeing), although we did not attempt to ``tune up'' the PSF ahead of time to reduce the testbed wavefront errors, which limit our detection sensitivity. The unocculted data is particularly unrealistic because we let the detector saturate to maintain the same exposure settings as the coronagraphic data. This is not typical for on-sky observations, and cannot be done at all with EM gain, sacrificing the benefits of the EM process. For high S/N observations with the coronagraph, the limiting brightness will be the satellite spots. So, for example, with \qty{50}{\nano\meter} astrogrid we improve our background-limited contrast by $10^{1.52}$ compared to the same data taken without the coronagraph with the exposure setting adjusted to avoid saturation. Without using the astrogrid, the long-exposure limit will be reached when the speckle halo saturates, which offers $\sim$\num{e2} improved contrast, which means VAMPIRES can observe targets $\sim$100 times brighter than its current limits.

\subsection{On-sky results}\label{sec:onsky}

We have not had any opportunities to use the coronagraph in good conditions for testing; most of our on-sky tests had $>$\ang{;;1.5} seeing. Our best data is from May 12, 2022, of target HIP 56083. We used a \qty{50}{\nano\meter} amplitude astrogrid with a separation of 15.5 $\lambda/D$, the same as our testbed data. An example coronagraphic image is shown in \autoref{fig:satellite-spots}b. We used 300 EM gain with 200 ms exposure times, median combining 100 frames for a total of \qty{20}{\second} of integration time. This caused significant blurring from seeing but was necessary to achieve a high signal-to-noise ratio (S/N). In the future, we want to investigate further the trade-off between S/N and sharpness from shorter exposure times on coronagraphic contrast.

\begin{figure*}
   \centering
   \includegraphics[width=0.9\textwidth]{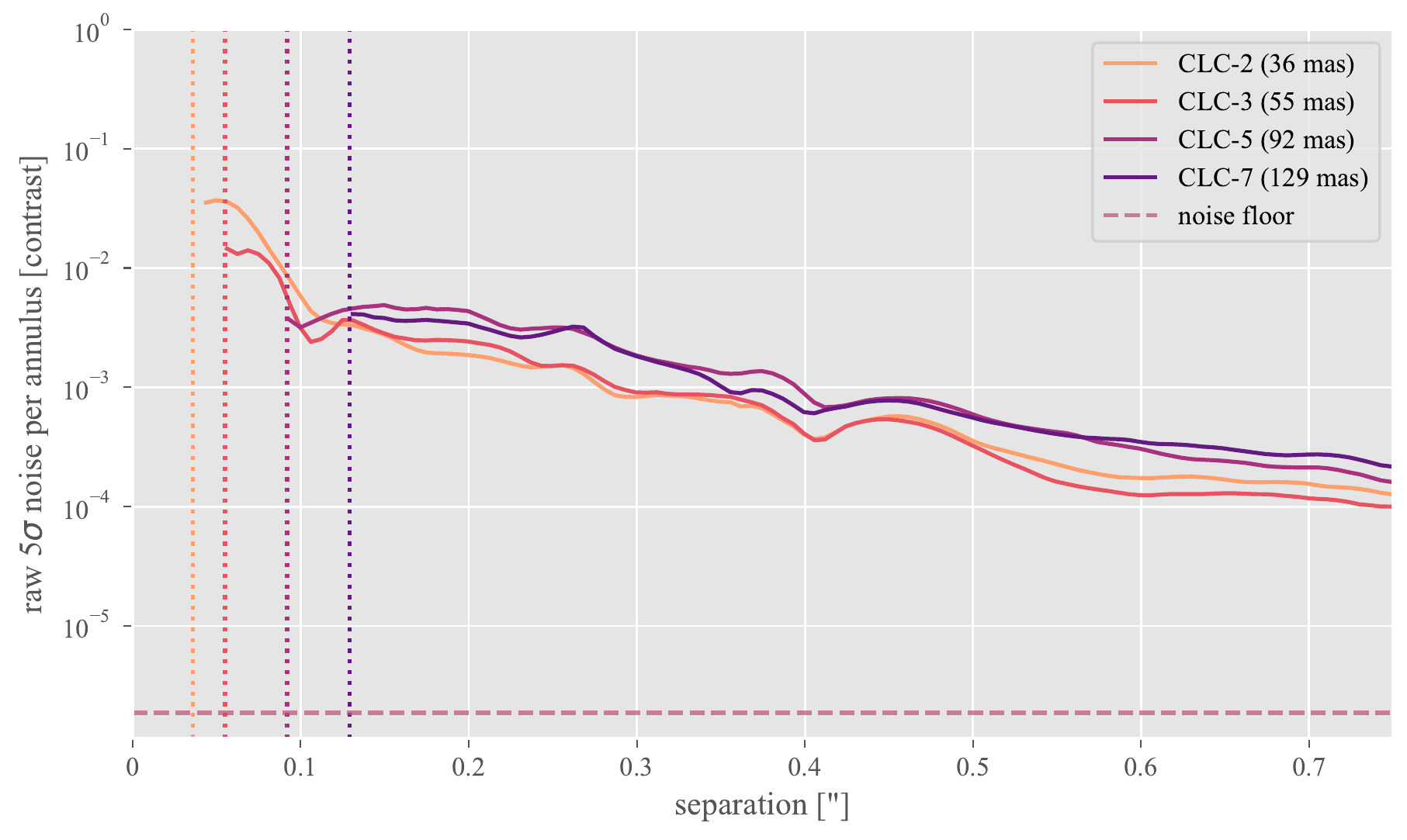}
   \caption{Raw 5$\sigma$ contrast curves measured from \qty{20}{\second} (\num{100} frames) of on-sky data of HIP 56083. These data were obtained on May 12, 2022, through the 750-50 filter with 300 EM gain and 200 ms exposures. The orange through purple curves correspond to each focal plane mask. The IWA for each curve is marked with a dotted vertical line. The astrogrid speckles are masked out.}\label{fig:onsky-contrast}
\end{figure*}

We constructed 5$\sigma$ contrast curves for each focal plane mask size following the same procedure as the testbed data (\autoref{fig:onsky-contrast}). Note for these observations we used a tighter field of view (256$\times$256) which limits the max separation to \ang{;;0.75}. We used the same scaling for the satellite spot photometry as fit from the testbed data ($10^{-1.52}$) and also masked out the satellite spots. At \ang{;;0.1}, all masks have a raw contrast around $10^{-2.5}$ and drop off to \num{e-4} at \ang{;;0.75}. The on-sky contrast is around an order of magnitude worse than the testbed data throughout the cropped \ang{;;0.75}$\times$\ang{;;0.75} FOV. We expect that under better observing conditions we could achieve many times better contrast performance with high S/N data aligned and stacked using lucky imaging for improved sharpness.

\section{Conclusions}\label{sec:conclusions}

In this paper, we have described the design, construction, and deployment of a visible-light Lyot coronagraph on SCExAO/VAMPIRES. This coronagraph attenuates $\sim$2 orders of magnitude of light, greatly improving the dynamic range of the EM-CCDs used on VAMPIRES with maximum EM gain. Additionally, the diffraction control will improve high-contrast performance, which has important applications in the optical for exoplanet science, both companion and protoplanetary disk imaging, as well as H$\alpha$ imaging. The coronagraph was approved for open-use observation starting in the 2022B observing semester.

Four partially transmissive focal plane masks were installed with IWAs of \qtylist{36;55;92;129}{\milliarcsecond}, respectively. The Lyot stop has a measured throughput of 65.7\%. During operation, the coronagraph uses ``astrogrid'' artificial speckles for photometric and astrometric calibration. We tested the coronagraph using the SCExAO internal source for characterizing the IWA, Lyot stop throughput, satellite spot relative flux, and measured contrast curves for each mask size using \qty{2}{\second} of median-combined data. The contrast was similar for each mask size, spanning \num{e-3} to \num{e-5} from \ang{;;0.1} to $>$\ang{;;0.8}, offering 0.2 to 1 orders of magnitude improvement over saturated non-coronagraphic images. We found the background limit for our detectors was $\sim$\num{3e-6} (5$\sigma$). We also measured the contrast for each mask size using \qty{20}{\second} of median-combined data of HIP 56083 in poor seeing conditions. This data is highly limited by the seeing halo, and achieves raw 5$\sigma$ contrast of \num{e-2} to \num{e-4} from \ang{;;0.1} to \ang{;;0.75}.

In the future, we want to implement various speckle control algorithms which have been demonstrated with SCExAO for improving post-coronagraphic contrast. The ``DrWHO'' focal plane wavefront sensing algorithm\cite{skaf2021} reduces the effects of non-common path aberrations (NCPAs) between the wavefront sensor and VAMPIRES by altering the wavefront sensor reference for convergence of the AO loop. DrWHO has been demonstrated on-sky\cite{skaf2022} and coronagraphically in the near-IR\cite{skaf2021}. Another technique for improving post-coronagraphic contrast is spatial linear dark field control (LDFC), which uses electric field conjugation to ``learn'' the NCPAs while on a bright reference star, and then removes NCPAs from our science target using the learned NCPAs. LDFC has been demonstrated with the near-IR coronagraphs on SCExAO\cite{ahn2022}.

In addition to implementing advanced wavefront control algorithms, there are hardware changes planned for the facility AO system (AO188\cite{minowa2010}) which will improve wavefront control on SCExAO\cite{lozi2022}. With this improved wavefront control, we want to explore phase-based focal plane masks such as the four-quadrant phase mask (4QPM)\cite{rouanFourQuadrantPhase2007}. Phase-based masks offer improved contrast over amplitude-based designs like the classic Lyot style, but are highly sensitive to pointing errors and thus require fine tip-tilt control for effective use\cite{huby2017}. We are also interested in exploring apodized Lyot stops to achieve similar or improved performance in the presence of aberrations without sacrificing as much throughput.


\appendix    

\section{Code and Data Availability}\label{sec:code}

The code used for simulating the coronagraph (\autoref{sec:design}), reducing and analyzing testbed (\autoref{sec:testbed}) and on-sky data (\autoref{sec:onsky}), and for producing the figures in this paper are all available under an open-source license in a GitHub repository (\href{https://github.com/mileslucas/vampires-coronagraph}{mileslucas/vampires-coronagraph}). This code makes significant use of the Julia programming language\cite{bezanson2017} along with the open-source packages \texttt{HCIPy}\cite{por2018}, \texttt{ADI.jl}\cite{lucas2020}, \texttt{numpy}\cite{harris2020}, and \texttt{astropy}\cite{astropycollaboration2013,astropycollaboration2018}. The CAD drawings for the focal plane masks and Lyot stops are also available in the online repository. Data is openly available upon request to accommodate the large data volumes. Further requests or questions about code or data are welcomed.

\acknowledgments

We wish to recognize and acknowledge the significant cultural role and reverence that the summit of Maunakea has always had within the indigenous Hawaiian community. We are grateful and thank the community for the privilege to conduct observations from this mountain. We would like to thank Paul Sumner for his useful advice during the construction of the Lyot stop and focal plane masks. We thank Jonathan Williams, Thayne Currie, and Timothy Brandt for generously offering observation time to use VAMPIRES for additional characterization of the coronagraph on-sky. This research was funded by the Heising-Simons Foundation through grant \#2020-1823. N.S. and V.D. acknowledge support from NASA (Grant \#80NSSC19K0336). This work is based on data collected at the Subaru Telescope, which is operated by the National Astronomical Observatory of Japan. The development of SCExAO is supported by the Japan Society for the Promotion of Science (Grant-in-Aid for Research \#23340051, \#26220704, \#23103002, \#19H00703, \#19H00695, and \#21H04998), the Subaru Telescope, the National Astronomical Observatory of Japan, the Astrobiology Center of the National Institutes of Natural Sciences, Japan, the Mt Cuba Foundation and the Heising-Simons Foundation.


\bibliography{report} 
\bibliographystyle{spiebib} 

\end{document}